\documentclass[twocolumn]{aastex63}
\hypersetup{linkcolor=cyan,citecolor=cyan,filecolor=cyan,urlcolor=blue}

\usepackage{graphicx}
\usepackage{amsmath}
\usepackage{xcolor}
\newcommand{\Chi}{\mbox{$\widetilde{\chi}\;$}}
\newcommand{\ExB}{\mbox{\rm{E$\times$B}}\;}
\shorttitle{Acceleration of ions in shocks}
\shortauthors{K. Stasiewicz \& B. Eliasson}

\begin{document}

\title{Ion acceleration to MeV by the  ExB wave mechanism in collisionless shocks}


\author[0000-0002-2872-5279]{Krzysztof Stasiewicz}
\email{krzy.stasiewicz@gmail.com}
\affiliation{Department of Physics and Astronomy, University of Zielona G\'ora, Poland }
\affiliation{Space Research Centre, Polish Academy of Sciences, Warsaw, Poland }
\author[0000-0001-6039-1574]{Bengt Eliasson}
\email{bengt.eliasson@strath.ac.uk}
\affiliation{SUPA, Department of Physics, University of Strathclyde, Glasgow, G4 0NG, United Kingdom}

\begin{abstract}
It is shown that ions can be accelerated  to MeV energy range in the direction perpendicular to the magnetic field  by the ExB mechanism  of electrostatic waves. The acceleration occurs in discrete steps of duration being a small fraction the gyroperiod and can explain   observations of ion energization to  10 keV at quasi-perpendicular shocks and to 100-1000 keV at quasi-parallel shocks. A general expression is provided for the maximum energy of ions accelerated in shocks of arbitrary configuration. The waves involved in the acceleration are related to three cross-field current-driven instabilities: the lower hybrid drift (LHD) instability induced by the density gradients in shocks and shocklets,  followed by the modified two-stream (MTS) and electron cyclotron drift (ECD) instabilities, induced by the ExB drift of electrons in the strong LHD wave electric field.  The ExB wave mechanism accelerates heavy ions to  energies proportional to the atomic mass number, which is consistent with satellite observations upstream of the bow shock and also with observations of  post-shocks in supernovae remnants.
\end{abstract}

\keywords{acceleration of particles --  shock waves -- solar wind -- turbulence -- bow shock}

\raggedbottom

\section{Introduction} \label{sec1}
When the solar wind plasma streaming with a speed of  400 km\,s$^{-1}$ and containing protons with kinetic energy of 1 keV and the thermal spread of  20 eV interacts with the Earth's quasi-perpendicular bow shock, the ion temperature increases by a factor of 10 across the shock, while the plasma flow slows down during the compression of the solar wind plasma and magnetic field.  The heating process is also associated with the appearance of energetic particles at energies  10 keV, which implies significant acceleration of a suprathermal population of the solar wind ions.  The electron temperature also undergoes a rapid increase by a factor of 10 across the shock.

On the other hand,  when  the interplanetary magnetic field is in the
quasi-parallel direction to the shock normal, an extended upstream foreshock region \citep{Greenstadt:1995,Eastwood:2005} is formed, containing ULF waves,  turbulence, non-linear structures and field-aligned beams. In addition to the electron and ion heating comparable to that occurring in quasi-perpendicular shocks, observations upstream of the quasi-parallel shocks show energetic ions accelerated to hundreds keV, indicating a three to four orders of magnitude increase of the kinetic energy.

The energetic ions observed in quasi-parallel shocks are traditionally believed to be energized in a diffusive shock acceleration process.
The key assumptions of this model are: (i)  the solar wind ions are preheated at the shock  and partially reflected upstream, (ii)  there are moving  barriers in the upstream region that reflect these particles back to the bow shock. After multiple bouncing between these barriers the particles gain energy through the Fermi acceleration mechanism  \citep{Fermi:1949,Bell:1978,Lee:1982,Burgess:2012,Otsuka:2018}.
Because the interplanetary shocks that could provide the upstream reflecting boundary are rare phenomena there has been a continuous search for other obstacles, such as for example foreshock transients, needed for the Fermi process to work at the bow shock. In a new attempt, \citet{Turner:2018} have suggested that hot flow anomalies \citep{Thomsen:1988,Liu:2016} observed occasionally in the solar wind  could make such upstream barriers, or traps where the energization occurs autogenously.

All mechanisms relying on the Fermi process require nonlocal  magnetic traps/mirrors, which are difficult to justify for energetic particles  observed on every satellite passage upstream of the quasi-parallel shock, viz., whenever the interplanetary magnetic field changes direction to quasi-parallel. Furthermore, the foreshock transients propagate in the upstream direction, against the solar wind, so they would contribute to the deceleration and not to the acceleration of the trapped particles.  Any acceleration relying on multiple bouncing would also require interaction times much longer than those implied by the observations.
Thus, a local process that does not require moving magnetic mirrors, or electrostatic field barriers, would be more suitable to explain ion acceleration at quasi-parallel shocks.

It has been recently shown \citep{Stasiewicz:2020a, Stasiewicz:2020c,Stasiewicz:2020d} that particle heating  and acceleration in collisionless shocks of arbitrary orientation are related to the wave electric fields of drift instabilities triggered  by shock compression of the plasma. It is a local process that can be summarized as follows:

{\em  Shock compressions of  the density N and  the magnetic field B }    $\rightarrow$ {\em  diamagnetic current}  $\rightarrow$ {\em lower hybrid drift (LHD) instability} $\rightarrow$  {\em electron \ExB drift} $\rightarrow$   {\em modified two-stream (MTS) and electron cyclotron drift (ECD)  instabilities}   $\rightarrow$ {\em heating: quasi-adiabatic ($\chi_j<1$), stochastic  ($\chi_j>1$), acceleration ($\chi_j \gg 1$)}.

 The stochastic heating and acceleration of particle species  with charge $q_j$ and mass $m_j$ ($j=e$ for electrons, $p$ for protons, $i$ for general ions) is controlled by the function

\begin{equation}
\chi_j(t,\mathbf{r})  = \frac{ m_j}{q_j B^2} {\rm div}(\mathbf{E}_\perp)  \label{eq1}
\end{equation}
that depends on the $m/q$ ratio and is also a measure of the charge non-neutrality. It is a generalization of the heating condition from earlier works of \citet{Karney:1979,McChesney:1987,Balikhin:1993,Vranjes:2010}, where the divergence is reduced to the directional gradient $\partial E_x/\partial x$.
 The particles are magnetized (adiabatic) for $|\chi_j|<1$,  demagnetized (subject to non-adiabatic heating) for $|\chi_j|\gtrsim1$, and selectively accelerated to high perpendicular velocities when $|\chi_j|\gg1$.

 The term 'stochastic' is here used in the sense of chaos theory for deterministic systems and does not involve random variables. At a certain threshold value of $|\chi_j|$,  particles with initially nearby states can have positive Lyapunov exponents and divergent trajectories. This happens for $|\chi_j|\gtrsim 1$ when the interacting waves have zero frequencies such as at shocks or low frequencies comparable to or below the cyclotron frequency, $f\lesssim f_{cj}$ \citep{McChesney:1987,Balikhin:1993,Stasiewicz:2000}. At higher wave frequencies $f\gg f_{cj}$ \citep{Karney:1979}, stochastic motion sets in for particles having velocities near the phase velocity, $v\gtrsim v_{ph}=\omega/k$ with a threshold value $E/B\gtrsim(f_{cj}/f)^{1/3}v_{ph}/4$ for stochastic motion, which can be written in dimensionless variables as $|\chi_j|\gtrsim \Omega^{2/3}/4$ with $\Omega=f/f_{cj}$ and $|\chi_j|=m_jkE/q_jB^2$. Wave frequencies near cyclotron harmonics \citep{Fukuyama:1977} can also lead to resonant acceleration of particles with $v\gtrsim v_{ph}$ to form high-velocity tails in the distribution function. Thus, at high frequencies we have the formation of an 'acceleration lane' indicated by a green line in Figure \ref{Ffigure1}.

 Previous simulations have shown that ions at perpendicular bow shocks are stochastically bulk heated with typical values of $\chi_p\sim 60$ produced by the electric fields of the lower hybrid drift instability.
Electrons can also be  heated stochastically  on electron cyclotron drift  waves. However, in most cases they
undergo a quasi-adiabatic heating process, $T_{e\parallel}\approx T_{e\perp} \propto B^{\alpha}$, where $\alpha=1/3-2/3$ \citep{Stasiewicz:2020c,Stasiewicz:2020d}.

The aim of this paper is to show that ions can be accelerated to  MeV energies by electrostatic waves in the frequency range from the proton gyrofrequency $f_{cp}$ to the multiples of the electron gyrofrequency $nf_{ce}$ associated with the three cross-field, current-driven LHD, MTS, and ECD instabilities mentioned above. The acceleration mechanism requires $\chi \gg 1$ and can increase velocity of some particles by the \ExB drift velocity due to the wave electric field, i.e., by the speed $\widetilde{V}_E=\widetilde{E}_\perp/B$ \citep{Sugihara:1979,Dawson:1983,Ohsawa:1985}. The \ExB wave mechanism is  related to the surfatron mechanism at shocks \citep{Sagdeev:1966,Katsouleas:1983,Zank:1996,Ucer:2001,Shapiro:2003}, which requires wide front of coherent waves  and acceleration is done after multiple ion reflections between the shock and the upstream region \citep{Zank:1996,Shapiro:2001}.
In contradistinction, the \ExB mechanism works on much shorter time-scales at a fraction of a cyclotron period and much shorter spatial scales to reach significant energies by the interaction with incoherent bursts of waves.
It is coupled to the stochastic condition (\ref{eq1}), which makes it possible to obtain significant acceleration of protons, $10\,\mathrm{eV}\rightarrow 200\,\mathrm{keV}$ on intermittent and bursty waves observed at shocks and in  the magnetosheath.

\section{Stochastic heating and acceleration} \label{sec2}
Stochastic heating and  acceleration of charged particles by electrostatic waves can be studied with the simulation setup used by \citet{Stasiewicz:2020c,Stasiewicz:2020d}.  In the  magnetic field $\mathbf{B}_0=(0,0,B_0)$ there is  a macroscopic convection electric field $E_{y0}$ that drives particles into an  electrostatic wave $\widetilde{E}_x=E_{x0}\sin( \omega_{D} t -k_x x )$ with wavenumber $k_x=2\pi/\lambda$, and the Doppler shifted frequency $\omega_{D}$  in the spacecraft frame.
Trajectories and velocities of particles  with mass $m$ and charge $q$  are  determined by the Lorentz equation $md\mathbf{v}/dt=q(\mathbf{E}+ \mathbf{v}\times \mathbf{B}_0)$.
By using dimensionless variables with time normalized by $\omega_c^{-1}$, space by $k_x^{-1}$ and velocity by $\omega_c/k_x$ with $\omega_c=q B_0/m$ being the angular cyclotron frequency, a system of  equations is obtained in the plasma reference frame

\begin{align}
   \frac{du_x}{dt}&= \Chi \sin(  \Omega t -x)+ u_y,\label{eqh1} \\
   \frac{du_y}{dt}&= -u_x,\label{eq3}\\
  \frac{d x}{dt}&=u_x,\qquad
  \frac{d y}{dt}=u_y,\label{eqh2}
\end{align}
that depends on two parameters: the normalized wave frequency in the plasma frame
$\Omega =(\omega_D -k_x E_{y0}/B_0)/ \omega_c $,
and

\begin{equation}
 \Chi  =\frac{E_{x0}}{B_0}\frac{k_x}{\omega_c},
 \end{equation}
the stochastic heating parameter (\ref{eq1}) for a single wave mode.  This is in fact the normalized amplitude of the wave induced \ExB drift speed $\widetilde{V}_{E,y}=\widetilde{E}_x/B_0$, not to be confused with the convection drift $V_{E,x}=-E_{y0}/B_0$ that is absorbed in the normalized frequency $\Omega$.
The initial gyration velocity of  a particle  is $(v_{x0},v_{y0},0)$, and  $v_{\perp 0}=(v_{x0}^2+v_{y0}^2)^{1/2}$. In normalized variables  it becomes

\begin{equation}
u_{\perp 0}=k_\perp v_{\perp 0}/\omega_c=k_\perp r_{c}
\end{equation}
with the initial Larmor radius $r_{c}=v_{\perp {0}}/\omega_c$, and $k_\perp=k_x$.

For a statistical description of the particles we follow the procedure outlined in previous works \citep{Stasiewicz:2020c,Stasiewicz:2020d}, and carry out a set of test particle simulations for  $M=1,000$ particles, which initially are Maxwell distributed in velocity and uniformly distributed in space. The initial conditions are described by a two-dimensional Maxwellian distribution function of velocity components perpendicular to the magnetic field, which in  the normalized variables  can be written as

\begin{equation}
 {F}=\frac{1}{2\pi u_{\perp0}^2}\exp\bigg(-\frac{(u_x^2+u_y^2)}{2 u_{\perp0}^2}\bigg).
\end{equation}
Here, $u_{\perp0}=v_{T0}k_\perp/\omega_c$ and the thermal speed $v_{T0}=(2T_0/m)^{1/2}$.

\begin{figure}
\includegraphics[width=\columnwidth]{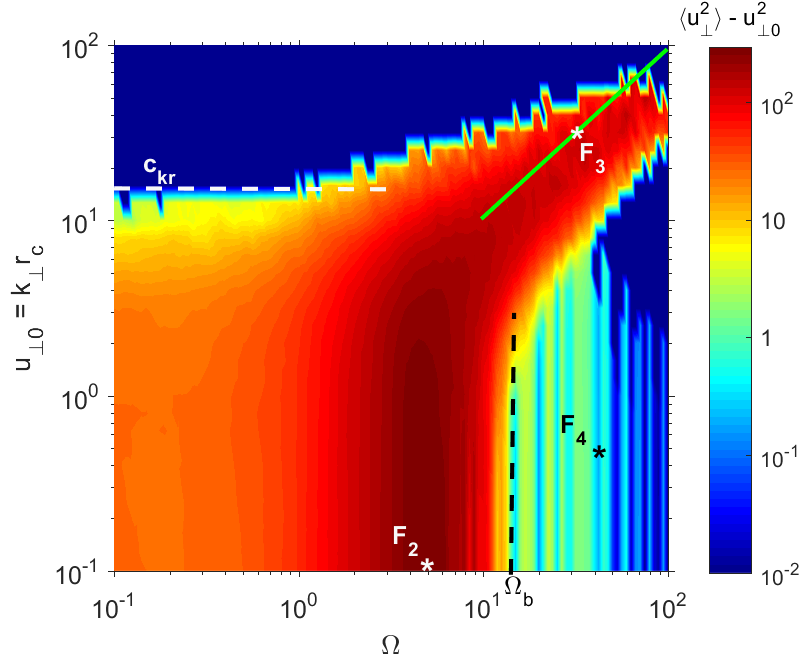}
\caption{A colormap of stochastic energization showing the difference $\langle u_\perp^2\rangle-u_{\perp0}^2$ between the normalized squared speed  at the end of the simulation and the initial value $u_{\perp0}^2=k_x^2 r_c^2$ after 1 cyclotron period for charged particles in an electrostatic wave with normalized electric field amplitude $\widetilde{\chi}=30$. Bulk heating takes place for $\Omega\lesssim \Omega_b\approx 13$ (vertical dashed line) and for $u_{\perp0}\lesssim c_{kr} \approx 15$, while for  $u_{\perp0}\gtrsim c_{kr}$, there is significant heating only for thermal velocity comparable to the wave phase velocity, or $u_{\perp0}\sim\Omega$ in the normalized variables (diagonal green line) leading to a distribution function having a high energy tail of particles. Positions marked with asterisk refer to Figures \ref{Ffigure2}, \ref{Ffigure3}, and \ref{Ffigure4} where the acceleration process is studied in detail.  \label{Ffigure1}}
\end{figure}

The system (\ref{eqh1})-(\ref{eqh2}) is advanced in time using a St{\"o}rmer-Verlet scheme \citep{Press:2007}. Simulations are carried out for several values of the normalized wave frequency $\Omega$ in the range $10^{-1}$ to $10^2$, and for the initial normalized thermal speed $u_{\perp0}$ spanning $10^{-1}$ to $10^2$. The normalized amplitude of the electrostatic wave is set to $\widetilde{\chi}=30$, which is typical for lower hybrid waves measured at the Earth's bow shock \citep{Stasiewicz:2020c}. The simulations are run for a relatively short time of one cyclotron period, motivated by the observations of rapid ion heating at the bow shock. The normalized mean squared speeds $\langle u_\perp^2\rangle=k_\perp^2 \langle v_\perp^2\rangle/\omega_c^2$ at the end of the simulations are calculated as

\begin{equation}
  \langle u_\perp^2\rangle=\frac{1}{M}\sum_{l=1}^M (u_{x,l}^2+u_{y,l}^2).
\end{equation}
Figure \ref{Ffigure1}  shows a color plot of the difference $\langle u_\perp^2\rangle-u_{\perp0}^2$ between the normalized squared speed $\langle u_\perp^2\rangle$ at the end of the simulation and the initial value $u_{\perp0}^2=k_\perp^2 r_c^2$.

It can be seen that the bulk heating region is most intense for frequencies $\Omega <\Omega_b \sim 13$. For protons this limit corresponds to frequencies one-third of the lower hybrid frequency ($\Omega_{lh}\approx 43$), and for wavelengths satisfying

\begin{equation}
k_\perp r_c < c_{kr} \sim 15 ; \qquad r_c \lesssim 2\lambda_\perp. \label{eqK}
\end{equation}
Thus, the stochastic acceleration of bulk plasma  disappears, when  the thermal particle gyroradius becomes larger than two wavelengths.  The  frequency  limit, $\Omega_b$,  and also $c_{kr}$ limit would shift to  larger values for maps computed with larger \Chi \citep{Stasiewicz:2020c}.

There is also a region of the acceleration of suprathermal particles from the tail of the distribution function that occurs along the green line $k_\perp r_c \approx \Omega$, for $\Omega \gtrsim 10$ as seen in Figure~\ref{Ffigure1}.
Particles on this line, hereafter referred to as the {\em acceleration lane}, have gyration speed $v_{\perp 0}$ that matches the phase speed of waves \citep{Fukuyama:1977,Karney:1979}

\begin{equation}
v_{\perp 0} \approx   r_c\omega_{c} = \omega/k_\perp  = f\lambda_\perp,  \label{VPER}
\end{equation}
which links  $v_\perp$ with electrostatic waves ($f,\lambda_\perp$) that can  accelerate these particles.
While the bulk heating  is done stochastically for all particles satisfying (\ref{eqK}),  the perpendicular  acceleration to high velocities along the acceleration lane (\ref{VPER}) is selective and requires some speed and phase matching.

\subsection{The physics of the E$\times$B wave heating}

\begin{figure}
\includegraphics[width=8cm]{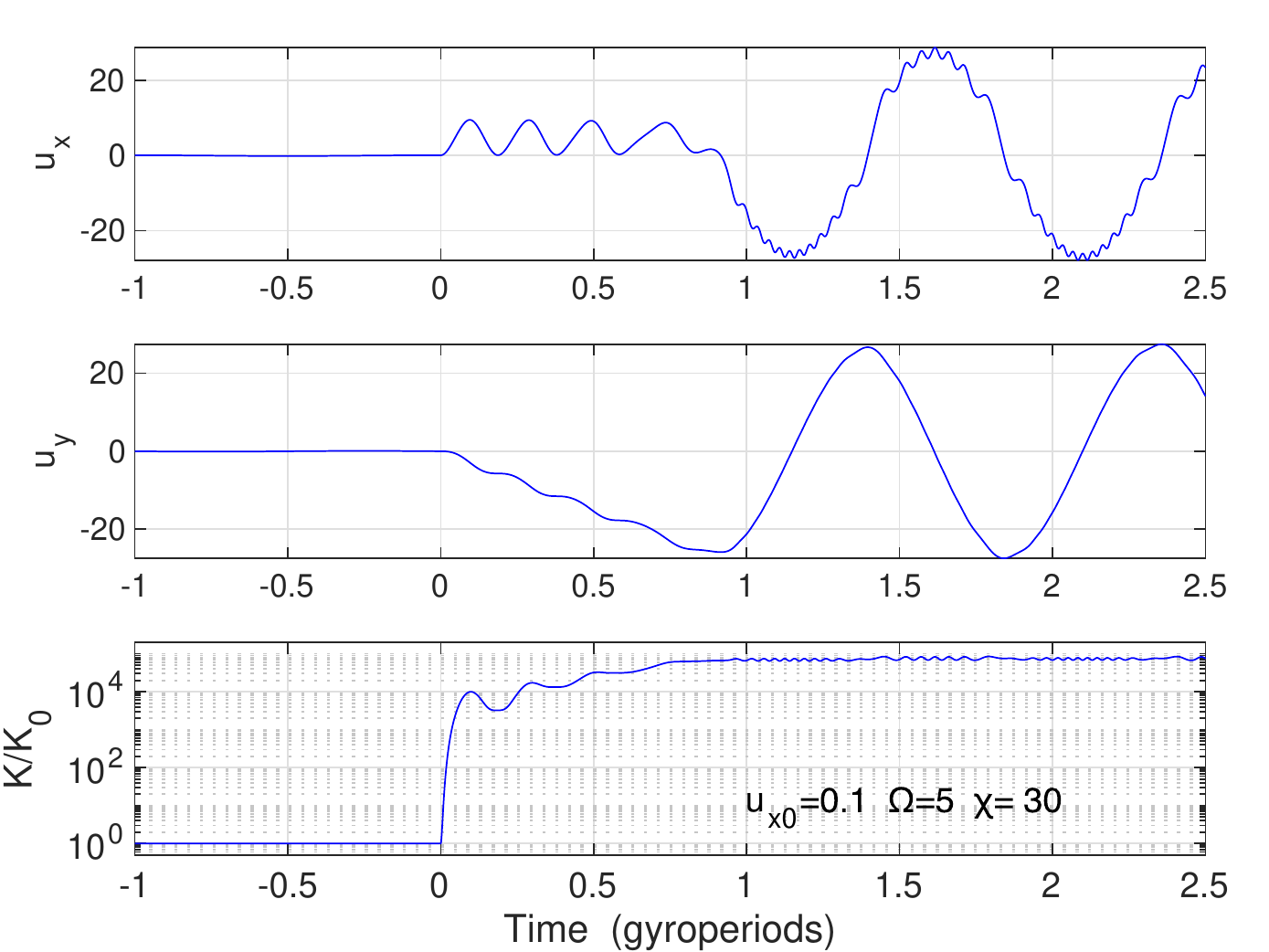}
\caption{Acceleration of a particle in the bulk heating region (marked as '$F_2$' in the colormap) with initial velocity $u_{x0}=k_x r_c=0.1,\; u_{y0}=0$ by wave $\Omega=5$ and $\Chi=30$ . The wave is switched on at $t=0$. Panels show velocity components $u_x$,  $u_y$, and the kinetic energy ratio $K/K_0$. Energy increase by a factor of   10$^4$ is achieved within half of the wave period, or 1/10 of the gyroperiod. \label{Ffigure2}}
\end{figure}

In order to understand the physics of the stochastic energization we have analyzed individual particle trajectories for cases  marked '$F_2$', '$F_3$', and '$F_4$' in Figure~\ref{Ffigure1}.
Figure~\ref{Ffigure2}  shows a solution of  Equations~(\ref{eqh1})-(\ref{eqh2}) for one  particle with  speed $u_{x0}=0.1$ and $u_{y0}=0$ injected into a wave at frequency $\Omega=5$ and amplitude $\Chi=30$ in the bulk heating region marked as '$F_2$' in Figure~\ref{Ffigure1}. The particle energy is increased by factor 10$^4$ within a half oscillation period of the electrostatic wave, corresponding to 1/10 gyroperiod.  In the beginning, the particle makes cyclotron motion with small velocity $u_\perp=0.1$ (not visible in the plot) until  $t=0$, when the wave is switched on. The velocity  $u_x(t)$  shows polarization drift response  $\propto d E_x/d t$, in the wave electric field,  before resuming the cyclotron motion after one gyroperiod. The velocity $u_y$ increases with time as the \ExB velocity $v_y(t) = - \widetilde{E}_x(t,x)/B_0$ to the maximum value in the normalized variables $u_{y} \approx (E_{x0}/B_0)(k_x/\omega_c)\equiv \Chi$.

\begin{figure}
\includegraphics[width=8cm]{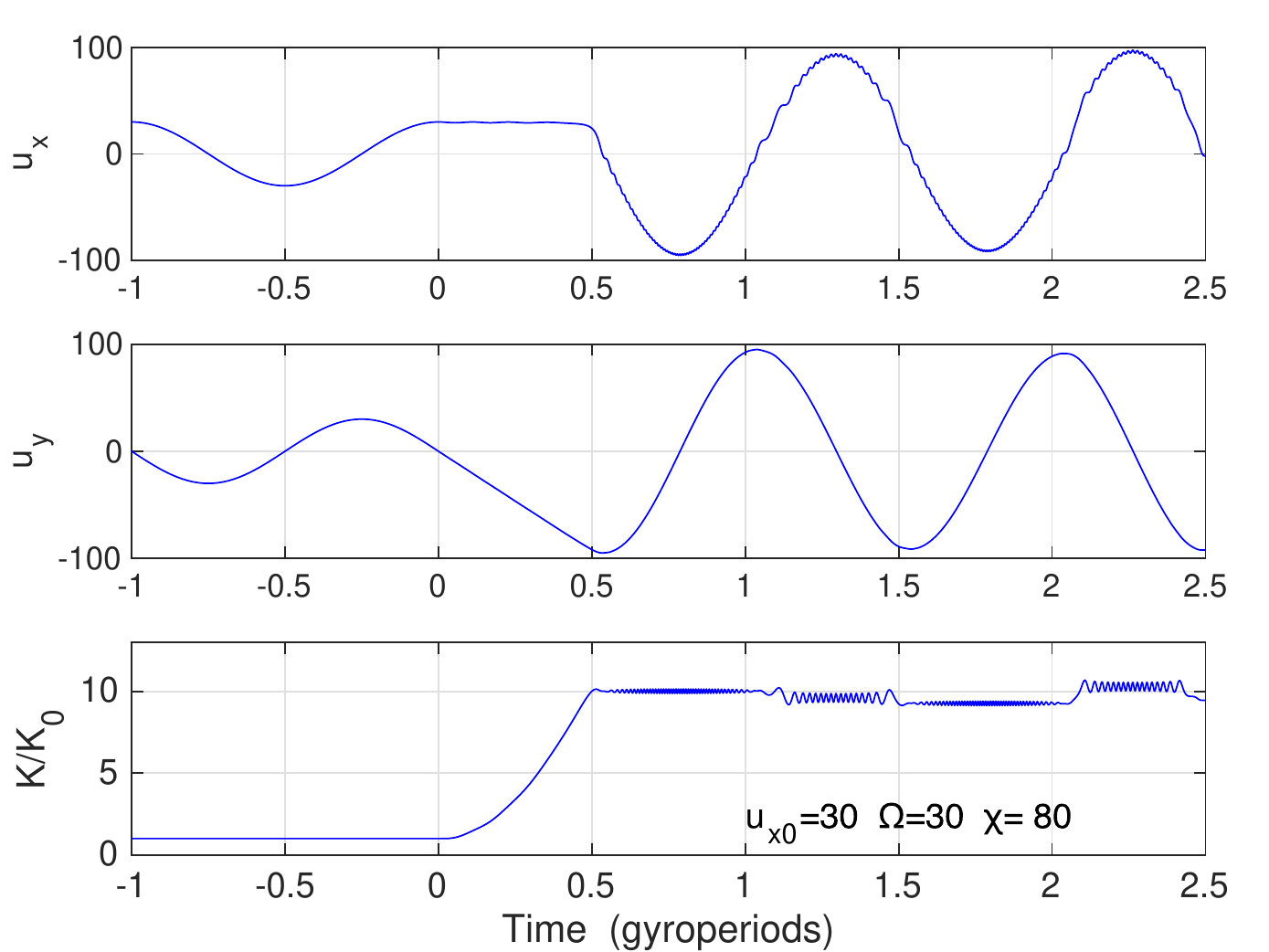}
\caption{The particle exiting Figure~\ref{Ffigure2} with  $u_{x0}=30$ encounters new wave $ \Omega=30,\; \Chi=80$ and is energized by a factor of 10 within  half of the gyroperiod. The position in the colormap is marked as '$F_3$'. \label{Ffigure3}}
\end{figure}

 \begin{figure}
\includegraphics[width=8cm]{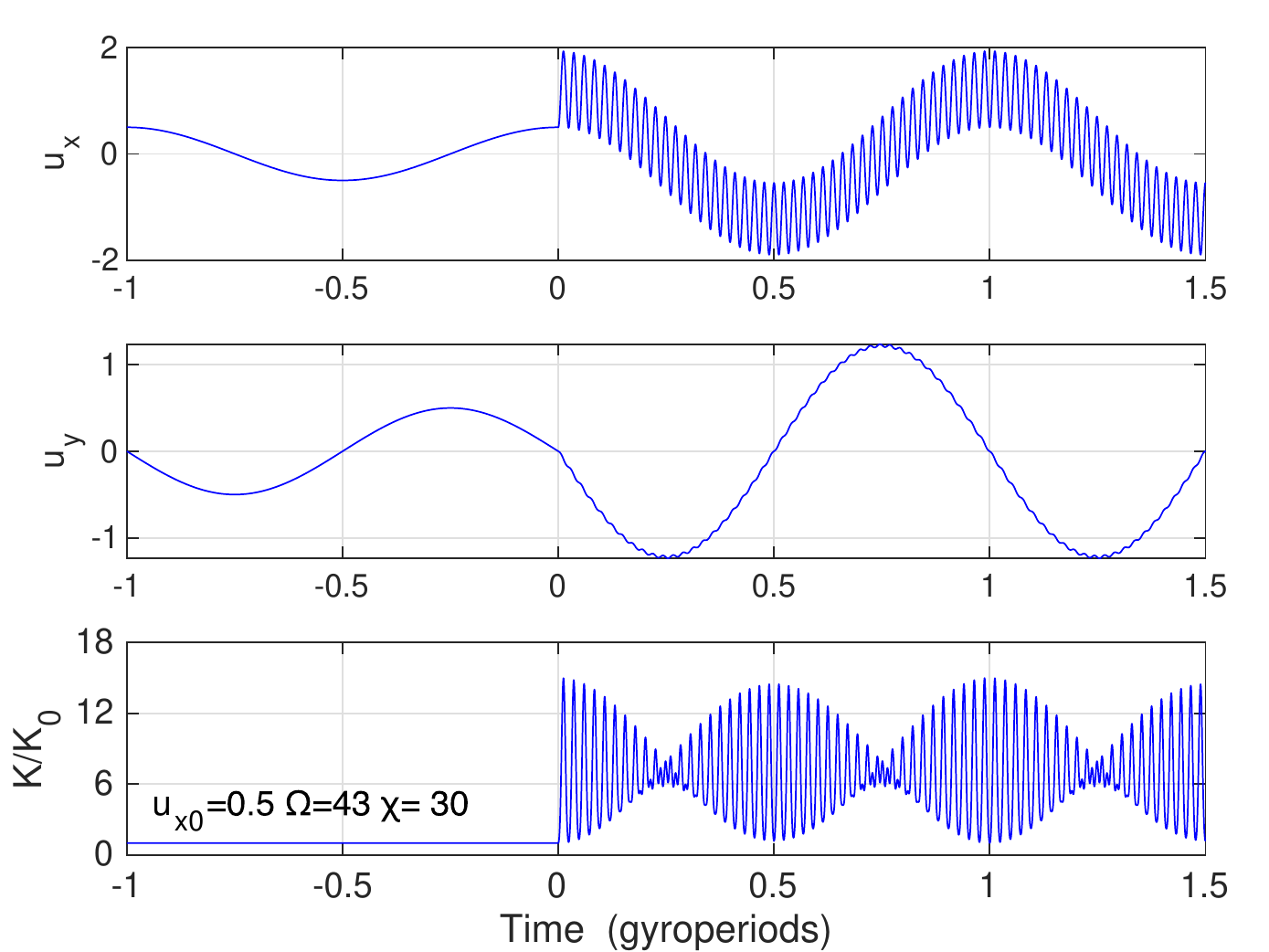}
\caption{Acceleration of a low energy particle with  $u_{x0}=0.5$ by a  lower hybrid wave $ \Omega=43,\; \Chi=30$. A proton is accelerated by a factor of 6 within 1/40 of the gyroperiod. The position in the colormap is marked as '$F_4$'. \label{Ffigure4}}
\end{figure}

The mechanism described above will be called '$\chi$-acceleration', or the  '\ExB acceleration', because
 the maximum acceleration capacity corresponds to the value of \Chi,  in normalized units, or to the \ExB velocity computed with the wave electric field, i.e., $\widetilde{V}_E=\widetilde{E}_\perp /B$ in physical units. This limiting value for the acceleration was previously found by \citet{Sugihara:1979} and \citet{Dawson:1983}, who analyzed the same equations~(\ref{eqh1})-(\ref{eqh2}) in the wave frame. This mechanism has been also used in simulations of ion heating by large amplitude magnetosonic waves by \citet{Lembege:1983}. The energization capacity  is then

\begin{equation}
K_{E}\lesssim \frac{m}{2} [v_{\perp 0}^2 + (\widetilde{E}_\perp/B)^2],    \label{KE}
\end{equation}
which is mass dependent. This equation is a general limit for the perpendicular acceleration of particles in quasi-parallel and quasi-perpendicular shocks as will be shown in section \ref{sec3}. It is applicable both to the bulk heating region, where $v_{\perp 0}=v_T$ is the thermal speed of particles, and also to the acceleration lane, where $v_{\perp 0}$ of suprathermal particles corresponds to the wave phase speed, or equivalently to $u_{\perp 0}=k_\perp r_c\sim \Omega$.

The acceleration capacity offered by equation (\ref{KE}) can be estimated from
 the  electric field measured on the Magnetospheric Multiscale (MMS) spacecraft \citep{Burch:2016} by  \citet{Ergun:2016,Lindqvist:2016}. In shocks the measured field    typically  exceeds 70 mV\,m$^{-1}$  for frequencies $\gtrsim 64$ Hz (available only in burst mode), with peak values of 300 mV\,m$^{-1}$. The magnetic field provided by \citet{Russell:2016}  can drop below 5 nT in the foreshock, so  the resulting velocity could be larger than $\widetilde{V}_E=\widetilde{E}/B \gtrsim 14,000$ km\,s$^{-1}$. With this value for speed we obtain  a minimum of 1 MeV as the capacity of the $\chi$-acceleration for protons at the bow shock.

The amplitudes of the wave electric field $\widetilde{E}$ and of \Chi  increase with frequency, which makes higher frequency waves more suitable for acceleration of particles to higher energies.
The lower frequency waves $\Omega<1$ ( $f<f_c=\omega_c/2\pi$) are inefficient accelerators because of  smaller amplitudes. They also require interaction times of a few cyclotron periods, but long coherent wave trains are  unlikely to occur in turbulent shock plasma.

\subsection{The acceleration lane and the polarization drift}

The particle accelerated to $u_x=30$ in the first step can encounter a new wave on the acceleration lane with frequency $\Omega=30$ and get additional energization as shown in Figure~\ref{Ffigure3}. The second wave with $\Chi=80$ would energize the particle  by a factor of   $10$ within a half gyroperiod. In this case $u_x(t)$ is constant, and $u_y(t)$ increases steady to the value of \Chi, i.e., to the \ExB speed in the wave field, until the cyclotron motion is resumed after $t=0.5$.  The second wave could be in any direction. The only requirement is that the phase speed of wave matches the perpendicular speed of a particle on an arbitrary phase of the gyration. The acceleration could continue along the acceleration lane, but it requires larger $\Omega$ and larger \Chi values on each subsequent step.
The acceleration works equally well for a conglomerate of waves with different frequencies and random phases \citep{Stasiewicz:2021b}.

By checking the effectiveness of the $\chi$-acceleration for different input parameters it is found that around  the acceleration lane ($u_{\perp 0}\sim \Omega$) the approximate energization rate is

\begin{equation}
  K/K_0 \sim 1+ (\Chi/u_{\perp 0})^2,  \label{KK0}
 \end{equation}
which  could continue to arbitrary high velocities $u_\perp$, providing there exist waves with sufficiently high amplitudes $\Chi \sim u_\perp$. The above expression is in fact equivalent to equation (\ref{KE}) derived in a  different way.

Yet another type of acceleration occurs for low energy particles in waves $\Omega>\Omega_b$, around the lower hybrid frequency $\Omega\approx 43$ (position '$F_4$' in Figure \ref{Ffigure1}) .  It is seen in Figure~\ref{Ffigure4} that a proton with velocity $u_{x0}=0.5$ is rapidly accelerated by an average factor of 6 within the wave period (1/40 of the cyclotron period), but it executes quivering motion related to the polarization drift seen in panel $u_x$. This means that the frequency $\Omega_b$ in Figure~\ref{Ffigure1} represents in fact the boundary between the  strong \ExB drift response for $\Omega<\Omega_b$, and a weaker polarization drift response for $\Omega>\Omega_b$.

Equation (\ref{VPER}) implies that particles with perpendicular energy $K_0$ and mass $m$ are on the acceleration lane when

\begin{equation}
K_0=\frac{m}{2}f^2\lambda_\perp ^2.   \label{EN}
\end{equation}
A handy formula for  ions with atomic mass $A=m_i/m_p$ is

\begin{equation}
K_{0\mathrm{[keV]}}\approx 10A f_\mathrm{[kHz]}^2 \lambda_{\perp \mathrm{[km]}}^2  \label{KEV}
\end{equation}
which applies also for electrons with $A=1/1836$.
Using this expression we can find, for example, that protons with  energy 1 keV could be  accelerated by waves $f=10$ Hz, $\lambda_\perp\approx 33$ km, which are in the lower hybrid range. On the other hand, protons at energy 1000 keV would interact  with waves $f=1$ kHz and $\lambda_\perp\approx 10$ km, which could be found in the ECD frequency range.
Oxygen ions  ($A=16$) at energy of 16 MeV would interact with the same  waves ($f\approx 1$ kHz and $\lambda_\perp \approx 10$ km) as 1 MeV protons.

The wave phase velocity $f\lambda_\perp =\omega/k_\perp$  in (\ref{EN})  determines the energy of particles prone to the acceleration by waves.
The LHD waves have maximum frequency $\omega_{lh}= (\omega_{ce}\omega_{ci})^{1/2}$ and wavenumbers $k_\perp (r_e r_i)^{1/2}\sim 1$, as shown by \citet{Daughton:2003} and \citet{Umeda:2018}, so the phase speed of LHD waves is $v_{ph}=f \lambda_\perp \sim  (v_{Te}v_{Ti})^{1/2}$.  Here, $v_{Te}=(2T_e/m_e)^{1/2}$ is the electron thermal speed, $v_{Ti}=(2T_i/m_i)^{1/2}$ is the ion thermal speed and the gyroradii are: $r_e=v_{Te}/\omega_{ce},\: r_i=v_{Ti}/\omega_{ci}$. This gives the maximum energy of particles accelerated by LHD waves with $\Omega \approx \omega_{lh}/\omega_{cp} \lesssim 43$ as

\begin{equation}
K^{\mathrm{LHD}} \lesssim 1.5 \left(\frac{m_i}{m_e} T_eT_i\right)^{1/2}, \label{LHD}
\end{equation}
where the factor 1.5 is an empirical factor that fits the energy of the accelerated ions in perpendicular shocks as shown in   section \ref{sec3}.  This value can be compared with factor of 2 implied by Equation (\ref{KK0})  when $\Chi  \sim u_{\perp 0} <43$.
For temperatures  $T_e\approx 40$ eV, \, $T_i\approx 400$ eV we obtain the proton energy $K \sim$ 8 keV, which is typically observed as the upper acceleration energy at quasi-perpendicular shocks.

\subsection{Comparison with other models}

The  processes described in sections 2.1 and 2.2 have some components in common with the surfatron mechanism introduced by \citet{Katsouleas:1983} for the relativistic acceleration of electrons in laser plasmas. The surfatron idea is based on work by \citet{Sagdeev:1966} and  has been elaborated further in many papers \citep{Zank:1996,Shapiro:2001,Ucer:2001,Shapiro:2003,Eliasson:2005}.
It has been also used to explain acceleration in shocks of supernova remnants \citep{McClements:2001} and acceleration of cosmic rays \citep{Kichigin:2013}.
Namely, particles can be trapped and transported in the potential well during extended time, which leads to the acceleration in the perpendicular direction until the resulting Lorentz force exceeds the electrostatic force of the wave, and the particle becomes un-trapped.

The surfing acceleration, as explained by \citet{Shapiro:2001,Shapiro:2003}, applies to quasi-perpendicular shocks, where electrostatic waves propagate in the sunward, $x$-direction, while the particles are accelerated in the $y$-direction, tangentially to the shock front. The acceleration is mainly by the dc convection electric field $E_{y0}$, and partly by the wave field $E_x$ for trapped particles. The surfatron mechanism requires wide front of coherent waves, with several ion gyroradii width in the $y$-direction and acceleration is done after multiple ion reflections between the shock and the upstream region \citep{Shapiro:2001}. The surfatron mechanism of \citet{Katsouleas:1983} offered 'unlimited acceleration', but because of practical impossibility to create wide front of coherent waves both in the laboratory plasma and at the turbulent bow shock, the  ideas of efficient surfatron acceleration have not been confirmed experimentally.  Another problem with surfing acceleration is that the wave electric field strengths are likely above the threshold for the modulational instability that leads to the breakup of the wave and eventually wave collapse. This would make turbulent field structures that destroys the phase trapping necessary for the surfatron mechanism.

In contradistinction to the cited models of surfing acceleration, the  \ExB wave mechanism does not require extended surfing because it is coupled with the stochastic condition (\ref{eq1}). For large $\chi$ values, energization by a factor 10$^4$ can be done within the wave period $f^{-1}$ as seen in Figure~\ref{Ffigure2}. It corresponds to 1/40  of the proton gyroperiod for lower hybrid waves in Figure~\ref{Ffigure4}.

The \ExB wave mechanism does not require wide wavefronts as the classical surfing acceleration \citep{Katsouleas:1983,Ucer:2001,Shapiro:2003}, and  the acceleration can be done by bursty intermittent wave packets as observed in satellite data shown in Figure~\ref{Ffigure6}. It has been demonstrated recently \citep{Stasiewicz:2021b} that a conglomerate of waves with a wide range of frequencies and random phases can accelerate protons from 10 eV to 100 keV within a gyroperiod. The proton energy flux obtained from simulations accurately reproduces the measured ion spectra at the bow shock.

The \ExB wave mechanism supported by Equation (\ref{eq1}) operates  not only at quasi-perpendicular and quasi-parallel shocks \citep{Stasiewicz:2020c,Stasiewicz:2020d}, but also, for example, in laboratory plasma during ion heating by drift waves \citep{McChesney:1987}, and in the ion heating regions of the topside ionosphere \citep{Stasiewicz:2000}.

Both shock surfing acceleration and shock drift acceleration \citep{Ball:2001} rely on macroscopic convection electric field to accelerate particles. The present mechanism uses only the wave electric field.  The wave amplitudes measured in shocks above the lower hybrid frequency  are typically 10-100 times larger than the convection field, which ensures rapid acceleration and high energization ratios. As will be shown later, it is most efficient in parallel shocks, where the average convection field is zero.

Other models require some pre-acceleration or heating, before they can be operational. The heating map in Figure~\ref{Ffigure1} can explain both, a rapid heating of 10 eV particles by a factor of 10$^4$, and further acceleration of 1 MeV ions along the acceleration lane. As mentioned earlier, the \ExB acceleration works within a fraction of the gyroperiod, while the shock surfing acceleration \citep{Zank:1996,Ucer:2001,Shapiro:2003} requires many cyclotron periods, and the diffusive shock acceleration \citep{Bell:1978,Lee:1982} requires even much longer times.

In the next section we show  measurements of waves and turbulence at quasi-perpendicular and quasi-parallel shocks, which indicate that these waves are likely to $\chi$-heat bulk of ions and also accelerate some particles to high energies by the \ExB mechanism presented above.

\begin{figure}[ht]
\includegraphics[width=\columnwidth]{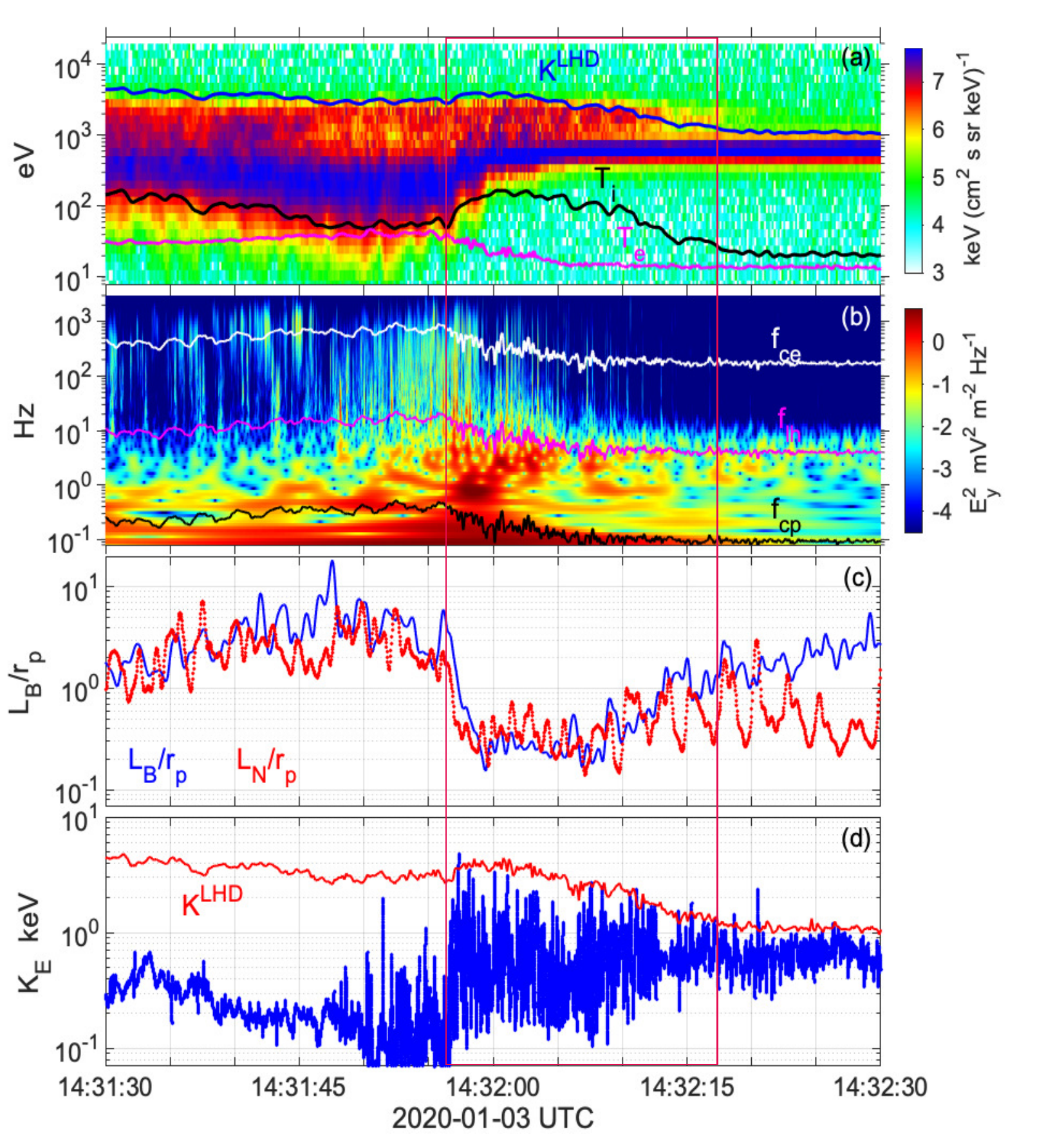}
\caption{Quasi-perpendicular shock measured by the MMS3 spacecraft. (a) Time versus energy spectrogram of the ion differential energy flux measured by FPI. Overplotted are the electron and ion temperatures and the acceleration capacity of LHD waves given by equation (\ref{LHD}). (b) Time versus frequency spectrogram of the  $E_y$ (GSE) component of the electric field.  Overplotted are the electron cyclotron $f_{ce}$, the lower hybrid $f_{lh}$, and the proton cyclotron  $f_{cp}$. (c) The measured gradient scale of the magnetic field $L_B$ and of the plasma density $L_N$ normalized with the thermal proton gyroradius $r_p$. (d) The energization capacity of waves given by (\ref{KE}) for waves $f<20$ Hz in the lower hybrid frequency range. Overplotted is the acceleration capacity given by equation (\ref{LHD}). Active energization and heating of ions occur within the red box. \label{Ffigure5}}
\end{figure}

 \begin{figure}[ht]
\includegraphics[width=\columnwidth]{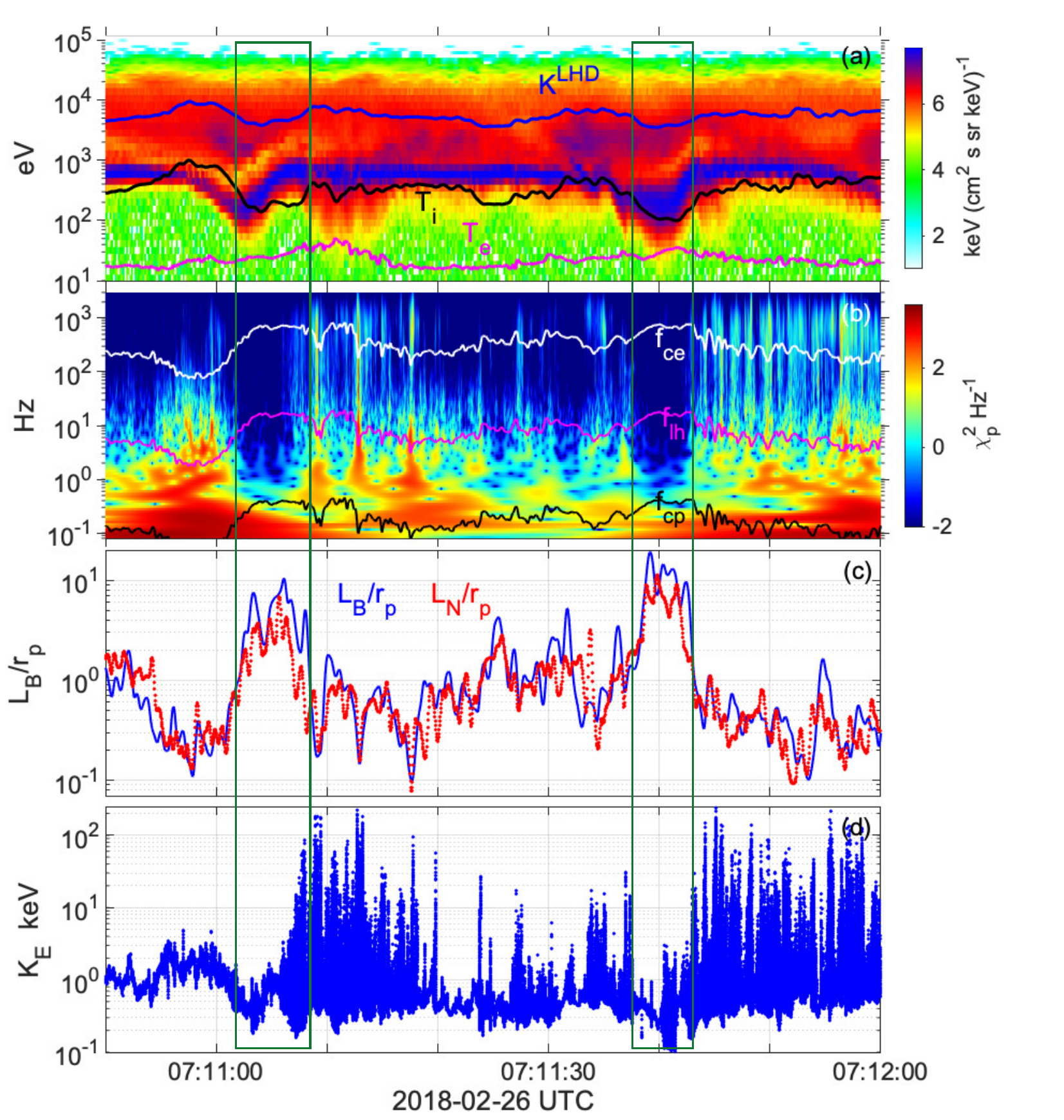}
\caption{Quasi-parallel shock measured by the MMS3 spacecraft.  (a) Ion energy flux spectrogram measured by FPI in the energy range 10 eV - 20 keV combined with EIS measurements in the energy range 20-100 keV.  Overplotted are electron and ion temperatures and the acceleration capacity of LHD waves given by equation (\ref{LHD}).  (b) Time versus frequency spectrogram of $\chi_p$. Overplotted are the plasma frequencies:  $f_{ce}$, $f_{lh}$, and   $f_{cp}$.  (c) Gradient scales $L_B$ and  $L_N$ derived from measurements and normalized with $r_p$. (d) The energization capacity of waves given by (\ref{KE}) for waves  in the measured frequency range 0-4000 Hz. Two green boxes mark two magnetic shocklets where the active ion energization does not occur.  \label{Ffigure6}}
\end{figure}

\section{Comparison with observations} \label{sec3}

Figure~\ref{Ffigure5} shows 1 minute of burst-mode data from the quasi-perpendicular bow shock. This is one of 9 multiple shock encounters analyzed by \citet{Stasiewicz:2020c}.   The particle data from the Fast Plasma Investigation  (FPI) \citep{Pollock:2016} shown in panel (a) are taken  at position (10.2, 13.4, -1.8) $R_E$ GSE (geocentric solar ecliptic).   The  Alfv\'{e}n Mach number was 7.2, the  electron plasma beta  $\beta_e\approx 1.1$, and the ion beta $\beta_i \approx 2.5$,  on the upstream (right) side of the shock. The angle between the magnetic field and the geocentric radial direction (a proxy to the shock normal) was 124$^\circ$. Overplotted are the ion and electron temperatures, and the acceleration capacity of LHD waves given by (\ref{LHD}). This equation provides accurate values for the maximum energy of protons accelerated at quasi-perpendicular shocks observed by MMS.

Active heating and acceleration of ions, seen in the ion temperature and the energy spectrum in panel (a) occur within the red box, which contains the ramp and the foot of the shock. In this region, ions are accelerated up to about 4 keV. The red box coincides with the region of the smallest values of the gradient scale lengths $L_B= B|\nabla B|^{-1}$ for the magnetic field and $L_N= N|\nabla N|^{-1}$ for the electron density $N$, both  normalized by the thermal ion gyroradius $r_p$ and shown in panel (c). The condition $L_N/r_p<(m_p/m_e)^{1/4}\approx 5$ determines the onset of the lower hybrid drift (LHD) instability \citep{Davidson:1977,Drake:1983,Gary:1993}, while $L_N/r_p<1$ in most of the time interval indicated by a red box in Figure \ref{Ffigure5}. The gradient scales are derived directly from  four point measurements using the method of  \citet{Harvey:1998}. It is seen that the values for  $L_N$ derived for the cold  solar wind, after 14:32:10 UTC are not reliable, and the values for $L_B$ should be used instead.

Almost the whole time interval in Figure \ref{Ffigure5} the plasma is unstable for the LHD instability, as seen in the wave spectrogram in panel (b) with the most intense waves in the frequency range $f_{cp}-f_{lh}$ located in the red box. These waves are indeed responsible for the  ion energization through the \ExB mechanism presented in section \ref{sec2}. This  can be seen in panel (d). The acceleration capacity of  waves below 20 Hz derived with (\ref{KE}) corresponds exactly to the limiting  energy of ions in panel (a), and coincides also with the other independent estimate (\ref{LHD}). The frequencies plotted in panel (b) are proportional to $B$ so the magnetic structure of the shock can be inferred from the frequency plots. Complementary discussion and overview of data for this case can be found elsewhere \citep{Stasiewicz:2020d}.

Figure~\ref{Ffigure6}  shows 1 minute of data from a long duration quasi-parallel shock measured by the MMS3 spacecraft.  The satellite was  at position (12.6, -3.9, 4.1) $R_E$, where the  Alfv\'{e}n Mach number was in the range 1-6 with the average of 3, the average electron plasma beta  $\beta_e\sim 0.7$, and the ion beta $\beta_i \sim 5$. The data represents  a couple of shocklets, i.e.,  compressions of the plasma density and of the magnetic field associated with  retardation of the solar wind beam as seen in panel (a). Two shocklets are marked with green boxes.  A major difference between this case and the previous one is that here ions are accelerated to up to about 100 keV, while in the quasi-perpendicular shock the ions were only accelerated to about 4 keV. In quasi-parallel shocks the acceleration of suprathermal particles extends well beyond the boundary $K^{LHD}$ as seen in panel (a).

The energization limit for  the measured waves computed with (\ref{KE})  is shown in panel (d). We see excellent agreement between the  theoretical maximum energy  $\sim 100$ keV in panel (d) and the measured energy spectra in panel (a).
The average gyroradius of a 40 keV proton in this time interval is 2000 km. Because of  large gyroradii of energetic ions, which tap energy from intermittent waves over large spatial areas, direct spatial correlations between $\sim \,$100 keV ions in panel (a) and accelerating waves in panel (d) are not expected. Such correlations do exist for low energy protons in Figure~\ref{Ffigure5}, in the red box.

The large difference in the maximum acceleration between quasi-perpendicular and quasi-parallel shocks appears to be related to the interaction time with waves. In perpendicular shocks, the solar wind is rapidly convected across the shock so the acceleration is done by LHD waves  up to the limit (\ref{LHD}), or to the limit (\ref{KE}) computed for lower hybrid waves only  ($f<20$ Hz), during a short time comparable to one gyroperiod. This observation indicates that the surfatron mechanism does not operate at quasi-perpendicular shocks. If the ions were reflected from the shock and remained longer time by surfing in the foot-ramp area they would have been accelerated to the limit (\ref{KE}), i.e. $\sim 100\,$keV also in quasi-perpendicular shocks, which is not  observed.

In parallel shocks, energetic  ions   meander between the shocklets in the upstream region and repetitively interact with higher frequency waves at increasing frequencies during much longer times. This would stepwise increase their energy to the limit (\ref{KE}) through the same $\chi$-acceleration mechanism, along the acceleration lane of Figure~\ref{Ffigure1}.

Let us analyze waves shown in panel (b).
The time versus frequency spectrogram of $\chi_p$ given by Equation~(\ref{eq1}) is derived from measurements of the electric field sampled at the rate 8192 s$^{-1}$.  The computed values reach  $\chi_p \approx 1800$ for higher frequency ECD waves. Details of the technique for computing $\mathrm{div}(\mathbf{E})$ from four point measurements are discussed by \citet{Stasiewicz:2020c,Stasiewicz:2020d}.

Figure 6(c) shows  $L_B/r_p$ and $L_N/r_p$ similar to Figure~\ref{Ffigure5}(c). Here, there is good agreement between the magnetic field and density length scales. The LHD waves in panel (b) are in  excellent correlation with regions $L_N/r_p \lesssim(m_p/m_e)^{1/4}\sim 5$, where the lower hybrid drift instability should theoretically occur.

As mentioned in section \ref{sec1}, the wave generation process in both cases is initiated by the   density gradients associated with the quasi-perpendicular shock in Figure~\ref{Ffigure5} and with quasi-parallel shocklets in Figure~\ref{Ffigure6}, which produce   diamagnetic currents that cause first the LHD instability \citep{Davidson:1977,Gary:1993,Daughton:2003} which has a lower threshold than the MTS and ECD instabilities.

 The wave spectrograms in Figures~\ref{Ffigure5}(b) and \ref{Ffigure6}(b) can be divided into four frequency bands: the magnetosonic waves below $f_{cp}$, the lower hybrid drift (LHD) waves in the frequency range $f_{cp}-f_{lh}$, the modified two-stream (MTS) instability in the range $f_{lh}-f_{ce}$, and the electron cyclotron drift (ECD) waves around and above $f_{ce}$. Other wave modes like whistlers and ion acoustic waves may also contribute in the spectrograms. The displayed spectrograms are in the spacecraft frame, so there may be some mixing and overlap of modes due to the frequency Doppler shift of short wavelengths by the bulk plasma flow $\sim 250$ km\,s$^{-1}$.

  In the frequency range $f_{cp}-f_{lh}$ there are  magnetic field fluctuations, which are also observed in simulations \citep{Daughton:2003}, in the magnetotail \citep{Ergun:2019}, and at the magnetopause \citep{Graham:2019}. This could mean that LHD waves coexist with ion  whistler waves created in the  density striations by mode conversion \citep{Rosenberg:2001,Eliasson:2008,Camporeale:2012} from LHD waves, or  with magnetosonic fluctuations. Such whistler waves, propagating upstream are seen in Figure~\ref{Ffigure5}.  Lower hybrid waves and whistlers can be also produced by ring distributions \citep{Winske:2015} of ions reflected from the bow shock, but Figure~\ref{Ffigure5}c and analysis of similar waves in Figure~\ref{Ffigure6} indicate that the driving mechanism  for LHD waves  at both shocks are density gradients rather than the reflected ion beams. However, the magnetosonic waves in the frequency range $f_{cp}-f_{lh}$ are equally efficient ion accelerators as demonstrated by \citet{Lembege:1983,Lembege:1984} and \citet{Ohsawa:1985}.

The enhanced electric field of the LHD or magnetosonic waves produces strong \ExB drifts of electrons only, because the ions are not subject to this drift due to their large gyroradius in comparison to the width of drift channels.
When the electron-ion  drift exceeds the ion thermal speed and becomes a significant fraction of the electron thermal speed, the MTS \citep{Wu:1983,Umeda:2014,Muschietti:2017} and ECD instabilities \citep{Lashmore:1973,Muschietti:2013,Janhunen:2018} are triggered at frequencies from above $f_{lh}$  to a few harmonics of $f_{ce}$. Such waves are commonly observed at the bow shock \citep{Wilson:2010,Breneman:2013,Goodrich:2018}. Note the vertical striations  in panels \ref{Ffigure5}(b) and \ref{Ffigure6}(b) that start from $\sim 0.5$ Hz (LHD instability) and extend up through the MTS and ECD instabilities  to 3 kHz, indicating co-location and common origin of these instabilities. The MTS waves propagate obliquely to the magnetic field and produce  parallel electric field component that may be responsible for the isotropisation  of the electron distribution  \citep{Stasiewicz:2020d}.

The sequential triggering and co-location of the LHD-MTS-ECD instabilities can be also explained by considering the expression for the \ExB drift velocity for particles with gyroradius $r_c$ in a spatially varying electric field $E_x\propto \sin(k_\perp x)$ \citep{Chen:2016}

\begin{equation}
\mathbf{V}_E= \frac{\mathbf{E}\times \mathbf{B}}{B^2}(1-\frac{1}{4}k_\perp^2r_c^2 ). \label{exb}
\end{equation}
Ions with large gyroradius  would have greatly reduced \ExB drift velocity in comparison with small gyroradius electrons. When the ratio $\lambda_\perp/r_p \lesssim \pi$, the ion electric drift vanishes, and the sole electron drift would produce strong cross-field current that could drive the above mentioned instabilities.
Actually, the conditions for the onset of the diamagnetic LHD instability on density gradients, and the complete quenching of the \ExB ion drift on short wavelengths are similar

\begin{equation}
\frac{L_N}{r_p} \sim \frac{\lambda_\perp}{r_p} \lesssim 5,
\end{equation}
which means that the chain of the instabilities LHD-MTS-ECD could be enforced by steepening of magnetosonic shock waves to smaller wavelengths, even in the absence of sufficient diamagnetic currents.

One should be also aware, that the \ExB drift of particles (\ref{exb}) is a  different phenomenon than the \ExB wave energization mechanism (\ref{KE}) discussed in this paper. The \ExB wave heating  of ions starts, when the  \ExB drift stops.

The ions accelerated by the $\chi$-mechanism in  quasi-parallel shocks  can diffuse through the magnetopause and form the quasi-trapped population of energetic ions inside. This idea is opposite to  claims that the energetic ions observed upstream of the bow shock represent leakage of particles from the magnetosphere \citep{Mauk:2019}.
 The dependence $\Chi \propto m/q$, and mass  dependence of the energization (\ref{KE},\ref{EN}) could explain observations that heavy ions in the C,N,O group have  fluxes larger than protons at high energies \citep{Stasiewicz:2013, Turner:2018}.
This is also consistent with observation of heavy ion temperatures $T_i\propto m_i/m_p$ in post-shocks of supernova remnants  \citep{Raymond:2017,Miceli:2019,Gedalin:2020}. However, there are also other explanations for the preferential heating of heavy ions \citep{Zank:1996,Zank:2001,Shapiro:2001}.

\section{Conclusions}
This research is based on the well established concepts of the stochastic heating  laid down in a seminal paper by \citet{Karney:1979}, represented by Equation (\ref{eq1}), and on the \ExB wave acceleration limit by large amplitude waves found by \citet{Sugihara:1979} and \citet{Dawson:1983}, represented by Equation (\ref{KE}). By combining these two concepts with multipoint MMS measurements  \citep{Burch:2016} we have shown  that  solar wind ions are bulk heated by the stochastic mechanism (\ref{eq1}) both in quasi-perpendicular and in quasi-parallel shocks confirming the previous results of \citet{Stasiewicz:2020c,Stasiewicz:2020d}. The perpendicular $\chi$-heating is a rapid process and may be accomplished within a fraction of a gyroperiod.  Selected  suprathermal particles with  perpendicular gyration velocity equal to the phase speed of electrostatic waves $v_\perp \approx  \omega/k_\perp$ can be  accelerated to velocities of the \ExB drift in the wave field,  $\widetilde{V}_E=\widetilde{E}_\perp/B$.
The acceleration  requires waves with the stochastic heating parameter $\Chi=(\widetilde{E}_\perp/B)(k_\perp/\omega_c) \gg 1$ and occurs in discrete steps on intermittent waves observed in shocks. The process  could bring  some ions  to the speed of $\sim 14,000$ km\,s$^{-1}$ or 1 MeV for protons, which is possible in quasi-parallel bow shocks where $\widetilde{E}\gtrsim 70$ mV\,m$^{-1}$ and $B\lesssim 5$ nT are observed. In the case analyzed in this paper protons are accelerated to $\sim 100$ keV and the theoretical prediction matches the measurements.

In collisionless shocks, waves that accelerate ions are produced by the three cross-field current-driven LHD, MTS, and ECD instabilities, in the frequency range $f_{cp}-nf_{ce}$, which are  seen in Figure~\ref{Ffigure6}(b). The instabilities are cascade-triggered by  diamagnetic currents induced by the density gradients created both  in  perpendicular  shocks and in shocklets that form parallel shocks.

The short interaction time with waves at perpendicular shocks limits the maximum energy of  protons accelerated by LHD waves to $\sim 10$ keV, while the multi-step acceleration by higher frequency waves $f_{lh}-nf_{ce}$ in parallel shocks can bring some ions to the  MeV energy range. The general expression (\ref{KE}) provides an explanation of the observed maximum energy of ions accelerated in shocks of arbitrary configuration.

 It is suggested that  ions accelerated in quasi-parallel shocks to hundreds keV diffuse into the magnetosphere and form the quasi-trapped energetic ion population.

The $\chi$- or \ExB-mechanism accelerates heavy ions to  energies proportional to the atomic mass number, which is consistent with satellite observations upstream of the bow shock and also with observations of ion temperatures in post-shocks of supernova remnants.


\acknowledgments
The data underlying this article are available to the public  through the MMS Science Data Center at
the Laboratory for Atmospheric and Space Physics (LASP),
University of Colorado, Boulder: https://lasp.colorado.edu/mms/sdc/public/.\\
The calibrated EIS data used in this paper were kindly provided by Ian J. Cohen at the Johns Hopkins APL.\\
 B.E. acknowledges support from the EPSRC (UK), grant EP/M009386/1.
\software{The  data were processed with the  IRFU-Matlab analysis package available at https://github.com/irfu/irfu-matlab.}

\bibliographystyle{aasjournal}

\end{document}